\documentclass[runningheads]{llncs}

\usepackage[T1]{fontenc}

\usepackage{graphicx}

\usepackage{amsmath}
\usepackage{mathtools}
\usepackage{amssymb}

\usepackage{hyperref}
\usepackage{color}

\usepackage[linesnumbered, ruled, vlined]{algorithm2e}

\usepackage{tikz}
\usetikzlibrary{positioning, arrows.meta}

\usepackage{listings}
\providecolor{dbluecolor}{rgb}{0.01,0.02,0.7}
\providecolor{dgreencolor}{rgb}{0.2,0.4,0.0}
\providecolor{dgraycolor}{rgb}{0.30,0.3,0.30}
\newcommand{\R}{\mathbb{R}}
\newcommand{\V}{\mathcal{V}}
\newcommand{\I}{\mathcal{I}}

\newcommand{\signs}{\{-, 0, +\}}
\newcommand{\func}[1]{\texttt{\textup{\detokenize{#1}}}}
\newcommand{\dd}[2]{\frac{\mathrm{d} #1}{\mathrm{d} #2}}
\newcommand{\abs}[1]{\lvert #1 \rvert}
\DeclareMathOperator{\support}{supp}
\newcommand{\supp}[1]{\support #1}
\DeclareMathOperator{\sign}{sign}
\DeclareMathOperator{\im}{im}
\DeclareMathOperator{\rank}{rank}
\DeclareMathOperator{\diag}{diag}

\newcommand{\len}{.6cm}
\newcommand{\runningS}{
	\begin{tikzpicture}[style={>={Stealth[round]}, semithick}]
		\node (1) {$A + B$};
		\node (2) [right=\len of 1] {$C$};
		\node (3) [below=\len of 2] {$D$};
		\node (4) [right=.3*\len of 2] {$A$};
		\node (5) [right=\len of 4] {$E$};
		\draw [->] ([yshift=1mm]1.east) to ([yshift=1mm]2.west);
		\draw [->] ([yshift=-1mm]2.west) to ([yshift=-1mm]1.east);
		\draw [->] (2.south) to (3.north);
		\draw [->] (3.north west) to ([xshift=2mm]1.south);
		\draw [->] ([yshift=1mm]4.east) to ([yshift=1mm]5.west);
		\draw [->] ([yshift=-1mm]5.west) to ([yshift=-1mm]4.east);
	\end{tikzpicture}
}
\newcommand{\runningK}{
	\begin{tikzpicture}[style={>={Stealth[round]}, semithick}]
		\node (1) {$a A + b B$};
		\node (2) [right=\len of 1] {$C$};
		\node (3) [below=\len of 2] {$c A + D$};
		\node (4) [right=.3*\len of 2] {$A$};
		\node (5) [right=\len of 4] {$E$};
		\draw [->] ([yshift=1mm]1.east) to ([yshift=1mm]2.west);
		\draw [->] ([yshift=-1mm]2.west) to ([yshift=-1mm]1.east);
		\draw [->] (2.south) to (3.north);
		\draw [->] (3.north west) to ([xshift=2mm]1.south);
		\draw [->] ([yshift=1mm]4.east) to ([yshift=1mm]5.west);
		\draw [->] ([yshift=-1mm]5.west) to ([yshift=-1mm]4.east);
	\end{tikzpicture}
}

\begin{document}

\title{A SageMath Package for\\
Elementary and Sign Vectors with\\
Applications to Chemical Reaction Networks}

\titlerunning{Package for Elementary and Sign Vectors with Applications to CRNs}

\author{Marcus S. Aichmayr\inst{1}\orcidID{0009-0008-8362-2644} \and
Stefan Müller\inst{2}\orcidID{0000-0002-3541-7856} \and
Georg Regensburger\inst{1}\orcidID{0000-0001-7735-3726}}

\authorrunning{M.~S.~Aichmayr et al.}

\institute{
	Institute of Mathematics, University of Kassel, Germany
	\email{aichmayr@mathematik.uni-kassel.de}\\
	\email{regensburger@mathematik.uni-kassel.de}
	\and
	Faculty of Mathematics, University of Vienna, Austria\\
	\email{st.mueller@univie.ac.at}
}

\maketitle

\begin{abstract}
    We present our \textsc{SageMath} package \texttt{elementary\_vectors} 
    for computing elementary and sign vectors of real subspaces.
    In this setting, elementary vectors are support-minimal vectors that can be determined
    from maximal minors of a real matrix representing a subspace.
    By applying the sign function,
    we obtain the cocircuits of the corresponding oriented matroid,
    which in turn allow the computation of all sign vectors
    of a real subspace.
    
    As an application,
    we discuss sign vector conditions
    for existence and uniqueness of complex-balanced equilibria
    of chemical reaction networks with generalized mass-action kinetics.
    The conditions are formulated in terms of sign vectors of two subspaces
    arising from the stoichiometric coefficients and the kinetic orders of the reactions.
    We discuss how these conditions can be checked algorithmically,
    and we demonstrate the functionality of our package
    \texttt{sign\_vector\_conditions} in several examples.

\end{abstract}

\let\thefootnote\relax\footnotetext{
	The \textsc{SageMath} packages are available at:
	\\
	\url{https://github.com/MarcusAichmayr/}
	\\
	The examples of this document are available at:
	\\
	\url{https://marcusaichmayr.github.io/sign_vector_conditions/}
}

\section{Elementary vectors of a subspace}

For real subspaces,
elementary vectors are nonzero vectors with minimal support,
as introduced in \cite{Rockafellar1969}.
Since we also deal with parameters,
we consider elementary vectors for subspaces $Q^n$
of the quotient field $Q$ of an integral domain $R$.
The \emph{support}\index{support} of a vector $x \in Q^n$
is the index set of nonzero components,
\begin{equation*}
	\supp x = \{i \mid x_i \neq 0\}.
\end{equation*}
\begin{definition}
	For a subspace $\V$ of $Q^n$,
	we call a nonzero vector $v \in \V$
	\emph{elementary}\index{elementary vector}
	if, for all nonzero vectors $x \in \V$,
	\begin{equation*}
		\supp x \subseteq \supp v
		\quad
		\text{implies}
		\quad
		\supp x = \supp v.
	\end{equation*}
\end{definition}
Elementary vectors with the same support are easily seen to be multiples.
Therefore, a subspace contains only finitely many elementary vectors up to multiples.
Further, a subspace is always generated by a finite set of elementary vectors.

We represent a subspace as the kernel of a matrix $M \in R^{d \times n}$ with rank~$d$
over~$Q$.
If the kernel of a matrix is $1$-dimensional, it has exactly one elementary vector.
We apply this fact by considering submatrices consisting of $d + 1$ columns.
To find kernel vectors, we solve systems of linear equations.
In particular, we compute maximal minors and apply Cramer's Rule.
By inserting additional zeros, we extend
the kernel vectors to kernel vectors of the initial matrix.
The resulting vectors are elementary if they are nonzero.

For $I \subseteq [n] = \{1, \ldots, n\}$,
we denote by $M_I$ the submatrix of $M$ that consists only of the columns
corresponding to the indices $I$.
We obtain a formula for computing elementary vectors in $R^n$.
\begin{proposition}[cf. equation (2.1) in \cite{Brualdi1995}]\label{prop:elementary_vectors}
	For a matrix $M \in R^{d \times n}$ with rank~$d$
	and $I \subseteq [n]$ with $\abs{I} = d + 1$,
	define the vector $v \in R^n$ by
	\begin{equation*}\label{eq:elementary_vectors}
		v_i
		=
		\begin{cases}
			(-1)^{\abs{\{k \in I \colon k < i\}}}
			\det M_{I \setminus \{i\}},
			& \text{if $i \in I$},
			\\
			0,
			& \text{otherwise}.
		\end{cases}
	\end{equation*}
	The vector $v \in \ker M$ is elementary if $\rank M_I = d$.
\end{proposition}
For computing all elementary vectors in $\ker M$, we need to compute $\binom{n}{d}$ determinants of all $d \times d$ submatrices of $M$ (maximal minors).
In contrast, for Gaussian Elimination, we would need to compute the kernel of
a $(d + 1) \times d$ matrix for each of the $\binom{n}{d + 1}$ index sets $I$.
In our implementation, it turns out that the approach using maximal minors is more efficient for computing elementary vectors for medium-sized matrices.

If all maximal minors are nonzero,
there is exactly one elementary vector for each index set $I$.
This gives an upper bound of $\binom{n}{d + 1}$ elementary vectors with pairwise distinct support
of an \mbox{$(n - d)$-dimensional} subspace.
If the rank of $M_I$ is not maximal, we obtain the zero vector.
Proposition~\ref{prop:elementary_vectors} suggests an algorithm for computing elementary vectors,
which we have implemented in our \textsc{SageMath} package \func{elementary_vectors} \cite{package_elementary_vectors}.
We demonstrate it by an example.
\begin{lstlisting}
	sage: from elementary_vectors import *
	sage: M = matrix([[1, 1, 2, 0], [0, 0, 1, 2]])
	sage: M.minors(2)
	[0, 1, 2, 1, 2, 4]
	sage: elementary_vectors(M)
	[(1, -1, 0, 0), (4, 0, -2, 1), (0, 4, -2, 1)]
\end{lstlisting}
Note that the first maximal minor is zero.
This is the reason why we obtain only $3$ and not $\binom{4}{3} = 4$ elementary vectors.

\subsubsection{Solvability of linear inequality systems.}\label{sec:solvability_of_linear_inequality_systems}
A fundamental theorem for deciding the solvability of linear inequality systems
is given in \cite{Rockafellar1969} and \cite{Rockafellar1970}.
Note that every linear inequality system can be written as an intersection of a subspace and a Cartesian product of intervals.
By iterating over elementary vectors,
we check whether such an intersection is empty.
For more details, we refer to our manuscript \cite{Aichmayr2024a}
(see also \cite{Minty1974}).
\begin{theorem}[``Minty's Lemma'', Theorem 22.6 in \cite{Rockafellar1970}]\label{thm:minty}
	For a subspace $\V$ of $\R^n$
	and nonempty intervals $\I = I_1 \times \cdots \times I_n$,
	\begin{itemize}
		\item[]
		either there exists a vector $x \in \V \cap \I$,
		\item[]
		or there exists an elementary vector $v \in \V^\perp$
		with $v^\top z > 0$ for all $z \in \I$.
	\end{itemize}
\end{theorem}
Based on Theorem~\ref{thm:minty},
the function \func{exists_vector}
decides the solvability of such systems.
\begin{lstlisting}
	sage: from vectors_in_intervals import *
	sage: M = matrix([[1, 0, 1, 0], [0, 1, 1, 1]])
	sage: I = intervals_from_bounds([2,5,0,-oo],[5,oo,8,5],
	....:      [True,True,False,False],[False,False,False,True])
	sage: I
	[[2, 5), [5, +oo), (0, 8), (-oo, 5]]
	sage: exists_vector(M, I)
	True
\end{lstlisting}

\subsubsection{Sign vectors.}\label{sec_sign_vectors}

We call elements in $\signs^n$ \emph{sign vectors}\index{sign vector}.
For $x \in \R^n$, we define the sign vector $\sign(x) \in \signs^n$ 
by applying the sign function componentwise.
For a set $S \subseteq \R^n$,
we obtain the set of sign vectors
\begin{equation*}
	\sign(S) = \{\sign(x) \mid x \in S\}
	\subseteq \signs^n
\end{equation*}	
by applying sign to each vector of $S$.
As for real vectors, we introduce the \emph{support}\index{support}
$\supp \sigma = \{i \mid \sigma_i \neq 0\}$
of a sign vector $\sigma$.

Further, we obtain a partial order on $\signs^n$ by defining $0 < -, +$.
As in~\cite{Mueller2019},
the \emph{(lower) closure}\index{lower closure} of a set of sign vectors
$T \subseteq \signs^n$ is the set
\begin{align*}
	\overline{T}
	=
	\{\sigma \in \signs^n \mid \text{$\sigma \leq \tau$ for some $\tau \in T$}\}.
\end{align*}
A (realizable) \emph{oriented matroid}\index{oriented matroid} is the set of sign vectors that correspond to a real subspace.
We call the elements of an oriented matroid \emph{covectors}\index{covector}.
To obtain them, we apply the sign function to the elements in a subspace.

The sign vectors corresponding to the elementary vectors are called \emph{cocircuits}\index{cocircuit}.
They generate all elements of an oriented matroid,
just like the elementary vectors generate all elements of a subspace.
Since the cocircuits are determined by the signs of the maximal minors, called \emph{chirotopes}\index{chirotope},
we can express many sign vector conditions in terms of chirotopes.
For further details on oriented matroids,
we refer to
\cite[Chapter 7]{Bachem1992},
\cite[Chapters 2 and 6]{Ziegler1995},
\cite{RichterGebert1997}
and the encyclopedic study~\cite{Bjoerner1999}.
Our package also offers several functions for oriented matroids.
As an example, we show the computation of cocircuits.
\begin{lstlisting}
	sage: from sign_vectors.oriented_matroids import *
	sage: M = matrix([[1, 1, 2, 0], [0, 0, 1, 2]])
	sage: cocircuits_from_matrix(M)
	{(-+00), (-0+-), (+0-+), (+-00), (0-+-), (0+-+)}
	sage: covectors_from_matrix(M)
	{(0000), (-+00), (--+-), (+-00), (+--+), (-0+-), (+0-+), (0-+-), (0+-+), (++-+), (-+-+), (+-+-), (-++-)}
\end{lstlisting}

\section{Applications to chemical reaction networks}

For chemical reaction networks (CRNs) with generalized mass-action kinetics, we recall basic notions from~\cite{Mueller2014}.
See also \cite{Mueller2019} and \cite{Mueller2012}.

A {\em generalized mass-action system} $(G_k,y,\widetilde y)$ is given by a simple directed graph $G=(V,E)$,
positive edge labels $k \in \R^E_>$,
and two maps ${y \colon V \to \R^n}$ and ${\widetilde y \colon V_s \to \R^n}$,
where $V_s = \{i \mid i \to i' \in E\} \subseteq V$ denotes the set of source vertices.
Every vertex $i \in V$
is labeled with a {\em (stoichiometric) complex} $y(i) \in \R^n$,
and every source vertex $i \in V_s$ is labeled with a {\em kinetic-order complex} $\widetilde y(i) \in \R^n$.
Further, every edge $(i \to i') \in E$ is labeled with a {\em rate constant} $k_{i \to i'} > 0$ and represents the {\em chemical reaction} ${y(i) \to y(i')}$.
If every component of $G$ is strongly connected, $G$ and $(G_k,y,\widetilde y)$ are called {\em weakly reversible}.
The associated ODE system for the positive {\em concentrations} $x \in \R^n_>$ (of $n$ chemical species) is given by
\begin{equation}\label{ode1}
	\dd{x}{t} 
	= \sum_{(i \to i') \in E} k_{i \to i'} \, x^{\widetilde y(i)} \big( y(i')-y(i) \big) .
\end{equation}
The sum ranges over all reactions, 
and every summand is a product of the {\em reaction rate} $k_{i \to i'} \, x^{\widetilde y(i)}$, involving a monomial $x^{\widetilde y} = \prod_{j=1}^n (x_j)^{\widetilde y_j}$ determined by the kinetic-order complex of the educt,
and the {\em reaction vector} $y(i')-y(i)$ given by the stoichiometric complexes of product and educt.

Let $I_E, I_{E,s} \in \R^{V \times E}$ be the incidence and source matrices
of the digraph~$G$, respectively,
and
\begin{equation*}
	A_k = I_E \diag(k) (I_{E,s})^\top \in \R^{V \times V}
\end{equation*}
be the Laplacian matrix of the labeled digraph~$G_k$.
(This definition is used in dynamical systems.
In other fields, the Laplacian matrix is defined as $A_k^\top$, $-A_k$, or $-A_k^\top$.)
Now, 
the right-hand-side of \eqref{ode1} can be decomposed into stoichiometric, graphical, and kinetic-order contributions,
\begin{equation}\label{ode2}
	\dd{x}{t}
	= Y I_E \diag(k) (I_{E,s})^\top \, x^{\widetilde Y}
	= Y A_k \, x^{\widetilde Y},
\end{equation}
where $Y,\, \widetilde Y \in \R^{n \times V}$ are the matrices of stoichiometric and kinetic-order complexes,
and $x^{\widetilde Y} \in \R^V_>$ denotes the vector of monomials,
that is, $(x^{\widetilde Y})_i = x^{\widetilde y(i)}$.
Clearly, the change over time lies in the {\em stoichiometric subspace}
$
S = \im (Y I_E),
$
that is, 
$\dd{x}{t} \in S$.
Equivalently, trajectories are confined to cosets of $S$,
that is, $x(t) \in x(0)+S$.
For positive $x' \in \R^n_>$, the set $(x'+S) \cap \R^n_>$ is called a {\em stoichiometric class}.

\noindent
The (stoichiometric) {\em deficiency} is given by
\begin{equation*}
	\delta 
	= \dim(\ker Y \cap \im I_E) 
	= |V| - \ell - \dim(S),
\end{equation*}
where $|V|$ is the number of vertices, 
and $\ell$ is the number of connected components of the digraph.
Analogously, we introduce the {\em kinetic-order subspace}
$
\widetilde S = \im (\widetilde Y I_E) 
$
and the {\em kinetic(-order) deficiency} 
$
\widetilde \delta
= \dim(\ker \widetilde Y \cap \im I_E)
= |V| - \ell - \dim(\widetilde S) .
$

A steady state $x \in \R^n_>$ of \eqref{ode2} that fulfills
$
A_k \, x^{\widetilde Y} = 0
$
is called a positive {\em complex-balanced equilibrium} (CBE).
On the one hand, if $\delta = 0$, then every equilibrium is complex-balanced.
On the other hand, if there exists a CBE,
then the underlying graph is weakly reversible, see the comment at the end of Section~4 in~\cite{Mueller2014},
and cf. \cite[Proposition~2.18]{Mueller2012}.

\begin{theorem}[robust $\delta = \widetilde{\delta} = 0$ theorem, Theorem 46 in \cite{Mueller2019}]
	\label{thm:robust_deficiency_zero_theorem}
	For a generalized mass-action system,
	there exists a unique positive CBE in every stoichiometric class,
	for all rate constants
	and for all small perturbations of the kinetic orders,
	if and only if $\delta = \widetilde{\delta} = 0$,
	the network is weakly reversible,
	and $\sign(S) \subseteq \overline{\sign(\widetilde{S})}$.
\end{theorem}
We consider a CRN
given by a graph with 5 vertices and 6 edges in 2 connected components,
and we label the vertices with stoichiometric and kinetic-order complexes, respectively. (The two resulting labeled graphs are shown below.)
The kinetic-order complexes involve parameters $a, b, c \in \R$.
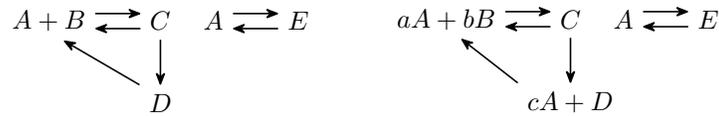
\begin{center}
	\runningS
	\qquad
	\runningK
\end{center}
The resulting stoichiometric and kinetic-order subspaces are given by
\begin{equation*}
	S
	=
	\im
	\begin{array}{cc}
	{\color{gray}
		\begin{array}{c}
			A \\
			B \\
			C \\
			D \\
			E
		\end{array}
		}
		& 
		\begin{pmatrix}
			-1 & 0 & -1 \\
			-1 & 0 & 0 \\
			1 & -1 & 0 \\
			0 & 1 & 0 \\
			0 & 0 & 1
		\end{pmatrix}
	\end{array}
	\quad
	\text{and}
	\quad
	\widetilde{S}
	=
	\im
	\begin{pmatrix}
		-a & c & -1 \\
		-b & 0 & 0 \\
		1 & -1 & 0 \\
		0 & 1 & 0 \\
		0 & 0 & 1
	\end{pmatrix}.
\end{equation*}
Clearly,
the network is weakly reversible, and it is easy to verify that
$\delta = \widetilde{\delta} = 0$.
To study existence and uniqueness of complex-balanced equilibria,
we compute the sign vectors of $S$ and $\widetilde{S}$.
\begin{lstlisting}
	sage: from sign_vectors.oriented_matroids import *
	sage: S = matrix([[-1,-1,1,0,0],[0,0,-1,1,0],[-1,0,0,0,1]])
	sage: covectors_from_matrix(S, kernel=False)
	{(00000), (00+-0), (0-0+-), (+000-), (+-++-), (0--+-), (+0+--), (+++--), (0-+--), (---+-), (--+0-), (--+--), (++0-0), ..., (++-++), (--+0+), (---++), (-+0-+)}
\end{lstlisting}
Since $\widetilde{S}$ depends on the parameters $a, b$ and $c$,
the sign vectors also depend on these parameters
and hence, we cannot compute them directly.
Thus, we consider specific values for the parameters and determine the corresponding sign vectors.
\begin{lstlisting}
	sage: var('a, b, c');
	sage: St = matrix([[-a,-b,1,0,0],[c,0,-1,1,0],[-1,0,0,0,1]])
	sage: covectors_from_matrix(St(a=2, b=1, c=1), kernel=False)
	{(00000), (+-++-), (0--+-), (+0+--), (+++--), (0-0+-), (0-+--), (---+-), (00-++), (+000-), (+0-+0), (--+--), (0++-0), ..., (++-+0), (-0+-0), (-+0-+), (-0+--)}
\end{lstlisting}
For $a = 2$, $b = 1$ and $c = 1$, $\sign(S) \subseteq \overline{\sign(\widetilde{S})}$.
Consequently, the conditions of Theorem~\ref{thm:robust_deficiency_zero_theorem} are satisfied in this specific case.
Obviously, we cannot cover all possible cases for the parameters that way.
However, by expressing this condition in terms of maximal minors of the kernel matrices,
we can compute with the parameters directly.
\begin{proposition}[Proposition 32 in \cite{Mueller2019}] \label{prop:closure_conditions}
	For subspaces $S$, $\widetilde{S} \subseteq \R^n$ of dimension $n - d$
	and matrices $W, \, \widetilde{W} \in \R^{d \times n}$
	with $S = \ker W$, $\widetilde{S} = \ker \widetilde{W}$,
	and rank~$d$,
	the following are equivalent:
	\begin{enumerate}
		\item $\sign(S) \subseteq \overline{\sign(\widetilde{S})}$.
		\item 
		$\det W_I \neq 0$ implies $\det W_I \det \widetilde{W}_I > 0$
		for all subsets $I \subseteq [n]$ with $\abs{I} = d$
		(or ``$< 0$'' for all $I$).
	\end{enumerate}
\end{proposition}
In the example, $S = \ker W$ and $\widetilde{S} = \ker \widetilde{W}$ with
\begin{align*}
	W
	=
	\begin{pmatrix}
		1 & 0 & 1 & 1 & 1 \\
		0 & 1 & 1 & 1 & 0
	\end{pmatrix}
	\qquad
	\text{and}
	\qquad
	\widetilde{W}
	=
	\begin{pmatrix}
		1 & 0 & a & a - c & 1 \\
		0 & 1 & b & b & 0
	\end{pmatrix}.
\end{align*}
By applying Proposition~\ref{prop:closure_conditions},
we obtain several conditions on $a$, $b$ and $c$.
We use package \cite{package_sign_vector_conditions}
to compute this (and several other) sign vector condition(s).
\begin{lstlisting}
	sage: W  = matrix([[1, 0, 1, 1, 1], [0, 1, 1, 1, 0]])
	sage: var('a, b, c');
	sage: Wt = matrix([[1, 0, a, a - c, 1], [0, 1, b, b, 0]])
	sage: from sign_vector_conditions import *
	sage: condition_closure_minors(W, Wt)
	[{a - c > 0, b > 0, a > 0}]
\end{lstlisting}
Hence, the network has a unique positive CBE
if and only if $a, b > 0$ and $a > c$.

\subsubsection{Uniqueness of CBE.} 

Also the uniqueness of CBE (in every stoichiometric class and for all rate constants) --- and hence its converse: multiple CBE -- can be characterized in terms of a sign vector condition.
For further details on sign vector conditions for injectivity in the context of CRNs and further references, we refer to \cite{Mueller2016a}.

\begin{proposition}[cf.\ Proposition 3.1 in \cite{Mueller2012}]
	For a generalized mass-action system, there exists at most one positive CBE
	in every stoichiometric class, for all rate constants, if and only if
	\begin{equation}\label{eq:intersection}
		\sign(S) \cap \sign(\widetilde{S}^\perp)
		=
		\{0\}.
	\end{equation}
\end{proposition}
Because of the parameters, we cannot directly compute the sign vectors in the kernel of $\widetilde{S}$.
Again, we use maximal minors to express \eqref{eq:intersection}.
\begin{corollary}[cf. Corollary~4 in \cite{Mueller2019}]
	For subspaces $S$, $\widetilde{S} \subseteq \R^n$ of dimension $n - d$
	and matrices $W, \, \widetilde{W} \in \R^{d \times n}$
	with $S = \ker W$, $\widetilde{S} = \ker \widetilde{W}$,
	and rank $d$,
	the following are equivalent:
	\begin{enumerate}
		\item $\sign(S) \cap \sign(\widetilde{S}^\perp) = \{0\}$.
		\item 
			Either $\det W_I \det \widetilde{W}_I \geq 0$ for all $I \subseteq [n]$ with $\abs{I} = d$, \\
			or $\det W_I \det \widetilde{W}_I \leq 0$ for all $I \subseteq [n]$ with $\abs{I} = d$.
	\end{enumerate}
\end{corollary}
Comparing the maximal minors yields:
\begin{lstlisting}
	sage: condition_uniqueness_minors(W, Wt)
	[{a - c >= 0, a >= 0, b >= 0}]
\end{lstlisting}
Hence, positive CBE are unique
if and only if $a, b \geq 0$ and $a \geq c$.

\subsubsection{Unique existence of CBE.}
We discuss a novel algorithm for checking a certain degeneracy condition for subspaces
that is part of a characterization of the unique existence of CBE.
The result also involves the set of nonnegative sign vectors
$T_\oplus = T \cap \{0, +\}^n$
of a set of sign vectors $T \subseteq \signs^n$.
\begin{theorem}[$\delta = \widetilde{\delta} = 0$ theorem, Theorem 45 in \cite{Mueller2019}] \label{thm:unique_existence}
	For a generalized mass-action system,
	there exists a unique positive CBE
	in every stoichiometric class, for all rate constants,
	if and only if $\delta = \widetilde{\delta} = 0$,
	the network is weakly reversible, and
	\begin{enumerate}
		\item\label{eq1:thm:one_equilibrium}
		$\sign(S) \cap \sign(\widetilde{S}^\perp) = \{0\}$;
		\item\label{eq2:thm:one_equilibrium}
		for all nonzero $\widetilde{\tau} \in \sign(\widetilde{S}^\perp)_\oplus$,
		there is a nonzero $\tau \in \sign(S^\perp)_\oplus$
		with $\tau \leq \widetilde{\tau}$; and
		\item\label{eq3:thm:one_equilibrium}
		$(S, \widetilde{S})$ is nondegenerate.
	\end{enumerate}
\end{theorem}
We discussed the first condition above.
To check the second condition, we use nonnegative cocircuits.
For the third condition,
we reformulate Definition~13 in~\cite{Mueller2019}, regarding the (non-)degeneracy of two subspaces.
\begin{definition}\label{def:degenerate}
	A pair $(S, \widetilde{S})$ of subspaces of $\R^n$ is called
	\emph{degenerate}\index{degenerate} if
	there exists $z \in \widetilde{S}^\perp$ with a positive component such that
	\begin{enumerate}
		\item \label{cond1_def:degenerate}
		for all $I_\lambda = \{i \in [n] \mid z_i = \lambda\}$ for some $\lambda > 0$,
		there exists $\pi \in \sign(S)_\oplus$ such that $\pi_i = +$ iff $i \in I_\lambda$; and
		\item \label{cond2_def:degenerate}
		for all nonzero $\tau \in \sign(S^\perp)_\oplus$,
		we have $\supp \tau \not\subseteq \supp z$.
	\end{enumerate}
\end{definition}
That is, the pair $(S, \widetilde{S})$ is degenerate
if there exists $z \in \widetilde{S}^\perp$
such that its equal positive components are covered by nonnegative sign vectors of $S$,
and
there is no nonzero, nonnegative sign vector in $S^\perp$
such that its support is contained in $\supp z$.

Since we simply cannot iterate over the subspace $\widetilde{S}^\perp$,
we have to find a different approach.
Instead, we consider sets of nonnegative covectors in $\sign(S)$.
Then, we use Theorem~\ref{thm:minty} (Minty's Lemma) to decide whether a vector $z$ exists
that has positive equal components on the support of each of these covectors.

We reformulate Definition~\ref{def:degenerate}
by demanding the existence of a set of nonnegative covectors that cover all equal positive components of a vector~$z$.
\begin{definition}
	A pair $(S, \widetilde{S})$ of subspaces in $\R^n$ is \emph{degenerate}\index{degenerate}
	if there exists a set of nonzero covectors $T \subseteq \sign(S)_\oplus$
	with disjoint support and a vector $z \in \widetilde{S}^\perp$
	such that
	\begin{enumerate}
		\item 
		\begin{enumerate}
			\item 
			for all $\pi \in T$,
			$\lambda_\pi = z_i = z_j > 0$ for all $i, j \in \supp \pi$,
			\item
			for all $i \notin \bigcup_{\pi \in T} \supp \pi$,
			$z_i \leq 0$, and
			\item\label{cond:degenerate_set_lambda_distinct}
			the $\lambda_\pi$'s are pairwise distinct;
		\end{enumerate}
		and
		\item\label{cond2:degenerate_set}
		for all nonzero $\tau \in \sign(S^\perp)_\oplus$,
		we have $\supp \tau \not\subseteq \supp z$.
	\end{enumerate}
\end{definition}
If two $\lambda_\pi$'s are equal, we compose the corresponding covectors
to cover equal components of $z$.
Therefore, Condition~\eqref{cond:degenerate_set_lambda_distinct} is redundant.
For the same reason,
the supports of the covectors do not need to be disjoint.
Since nonnegative covectors can be represented as a composition of nonnegative cocircuits,
it suffices to consider cocircuits instead of covectors.
Further, note that we can check Condition~\ref{cond2:degenerate_set} using cocircuits.
Following these observations, we obtain another reformulation of Definition~\ref{def:degenerate}.
\begin{definition}
	A pair $(S, \widetilde{S})$ of subspaces in $\R^n$ is \emph{degenerate}\index{degenerate}
	if there exists a set of nonnegative cocircuits $C \subseteq \sign(S)_\oplus$ and a vector $z \in \widetilde{S}^\perp$
	such that
	\begin{enumerate}
		\item 
		\begin{enumerate}
			\item 
			for all $\pi \in C$,
			$z_i = z_j > 0$ for all $i, j \in \supp \pi$,
			and
			\item
			for all $i \notin \bigcup_{\pi \in C} \supp \pi$,
			$z_i \leq 0$;
		\end{enumerate}
		and
		\item 
		for all cocircuit $\tau \in \sign(S^\perp)_\oplus$,
		we have $\supp \tau \not\subseteq \supp z$.
	\end{enumerate}
\end{definition}
To check degeneracy algorithmically,
we iterate over sets of nonnegative cocircuits.
In particular, Algorithm~\ref{alg:degenerate} below recursively constructs such sets
and determines whether a corresponding $z$ exists.
One could store the cocircuits and construct a subspace of $\widetilde{S}^\perp$
using the conditions on equal components.
Here, we modify the subspace such that its elements are equal on the components corresponding to the cocircuits in each step (line~\ref{line_alg:subspace_equal_comp})
and keep track of the positive entries using an index set.
If this subspace contains a vector that is positive on exactly this index set
(line~\ref{line_alg:exists1}),
condition~\ref{cond1_def:degenerate} holds.
If this vector also satisfies condition~\ref{cond2_def:degenerate} (line~\ref{line_alg:degenerate_cond2}),
it certifies degeneracy.
For efficiency, we use cocircuits to check this condition.
Note that we apply \func{exists_vector}
from Section~\ref{sec:solvability_of_linear_inequality_systems}
to efficiently check for existence
in line~\ref{line_alg:exists1} and \ref{line_alg:exists2}.

\newpage
\begin{algorithm}[ht]
	\caption{\texttt{subspaces degenerate}}
	\label{alg:degenerate}
	\DontPrintSemicolon
	\SetKwFunction{degenerate}{degenerate}
	\SetKwFunction{rec}{recursive}
	\SetKwProg{Fn}{Function}{:}{}
	\Fn{\degenerate{$S, \widetilde{S}$}}{
		$C :=$ set of nonnegative cocircuits of $\sign(S)$\;\label{line_alg:nn_cocircuits}
		$\V := \widetilde{S}^\perp$\;
		\label{line_alg:vector_space}
		\func{global} $\func{is_degenerate} := \func{False}$\;\label{line_alg:global_variable}
		\rec{$C, \V, \emptyset$}\;
		\KwRet $\func{is_degenerate}$\;
	}
	\Fn{\rec{$C, \V, I$}}{
		\While{$C \neq \emptyset$\label{line_alg:loop}}
		{
			choose any $\pi \in C$\;
			$C := C \setminus \{\pi\}$\;
			$\overline{\V} :=$ subspace of $\V$
			where vectors are equal on $\supp \pi$\;
			\label{line_alg:subspace_equal_comp}
			$\overline{I} := I \cup \supp \pi$\;
			\uIf{$z \in \overline{\V}$ exists
			with $z_i > 0$ iff $i \in \overline{I}$\label{line_alg:exists1}}
			{
				\For{$\sigma \in \sign(\overline{\V})$
				with $\sigma_i = +$ iff $i \in \overline{I}$}
				{
					\uIf{
						$\supp \tau \not\subseteq \supp \sigma$
						for all cocircuits $\tau \in \sign(S^\perp)_\oplus$%
						\label{line_alg:degenerate_cond2}
					}
					{
						$\func{is_degenerate} := \func{True}$\;\label{line_alg:degenerate_certified}
						\KwRet\;
					}
				}
			}
			\uElseIf{$z \in \overline{\V}$ exists
			with $z_i > 0$ if $i \in \overline{I}$\label{line_alg:exists2}}
			{
				\rec{$C, \overline{\V}, \overline{I}$}\;
			}
			\lIf{$\func{is_degenerate}$}{\KwRet}\label{line_alg:break}
		}
		\KwRet\;
	}
\end{algorithm}

\noindent
We consider Example~20 from \cite{Mueller2019}.
Here, we have matrices
\begin{align*}
	W =
	\begin{pmatrix}
		0 & 0 & 1 & 1 & -1 & 0 \\
		1 & -1 & 0 & 0 & 0 & -1 \\
		0 & 0 & 1 & -1 & 0 & 0
	\end{pmatrix}
	\quad
	\text{and}
	\quad
	\widetilde{W} =
	\begin{pmatrix}
		1 & 1 & 0 & 0 & -1 & a \\
		1 & -1 & 0 & 0 & 0 & 0 \\
		0 & 0 & 1 & -1 & 0 & 0
	\end{pmatrix}.
\end{align*}
The existence of a unique positive CBE depends on $a > 0$.
\begin{lstlisting}
	sage: var('a'); assume(a > 0);
	sage: W=matrix(3,6,[0,0,1,1,-1,0,1,-1,0,0,0,-1,0,0,1,-1,0,0])
	sage: Wt=matrix(3,6,[1,1,0,0,-1,a,1,-1,0,0,0,0,0,0,1,-1,0,0])
\end{lstlisting}
The first two conditions of Theorem~\ref{thm:unique_existence}
are independent of $a$.
\begin{lstlisting}
	sage: condition_uniqueness_sign_vectors(W, Wt)
	True
	sage: condition_faces(W, Wt)
	True
\end{lstlisting}
Condition~\ref{eq3:thm:one_equilibrium}
holds iff $a \in (0, 1) \cup (1, 2)$
as we demonstrate for specific values.
\begin{lstlisting}
	sage: condition_nondegenerate(W, Wt(a=1/2))
	True
	sage: condition_nondegenerate(W, Wt(a=2))
	False
\end{lstlisting}

\begin{credits}
	\subsubsection{\ackname} 
	This research was funded
	in part by the Austrian Science Fund (FWF), 
	grant 10.55776/P33218 to SM.

	\subsubsection{\discintname}
	The authors have no competing interests to declare.
\end{credits}

\end{document}